# Relationship Maintenance in Software Language Repositories

Ralf Lämmel[a]

a     Software Languages Team, Faculty of Computer Science, University of Koblenz-Landau, Germany

**Abstract**     The *context* of this research is testing and building software systems and, specifically, software language repositories (SLRs), i.e., repositories with components for language processing (interpreters, translators, analyzers, transformers, pretty printers, etc.). SLRs are typically set up for developing and using metaprogramming systems, language workbenches, language definition frameworks, executable semantic frameworks, and modeling frameworks.

    This work is an *inquiry* into testing and building SLRs in a manner that the repository is seen as a collection of language-typed artifacts being related by the applications of language-typed functions or relations which serve language processing. The notion of language is used in a broad sense to include text-, tree-, and graph-based languages as well as representations based on interchange formats and also proprietary formats for serialization.

    The overall *approach* underlying this research is one of language design driven by work on a particular SLR, YAS, which features a significant number of processed languages and language processors as well as a noteworthy heterogeneity in terms of representation types and implementation languages.

    The *knowledge* gained by our research is best understood as a declarative language design for regression testing and build management; we introduce a corresponding language Ueber with an executable semantics which maintains relationships between language-typed artifacts in an SLR.

    The *grounding* of the reported research is based on the executable, logic programming-based definition of the Ueber language and its systematic application to the management of YAS which consists of hundreds of language definition and processing components (such as interpreters and transformations) for more than thirty languages (not counting different representation types) with Prolog, Haskell, Java, and Python being used as implementation languages.

    The *importance* of this work stems from its objective of helping to understand and maintain relationships in SLRs, thereby directly helping users and developers of SLRs.

# The Art, Science, and Engineering of Programming

**Relationship Maintenance in Software Language Repositories**

# 1 Introduction

Build management including regression testing is an important current topic in software engineering, see, e.g., [43, 22, 32, 24] for some recent research. In this paper, we are specifically concerned with testing and building software language repositories (SLRs)—these are repositories with components for language processing (interpreters, translators, analyzers, transformers, pretty printers, etc.). SLRs are typically set up for developing and using metaprogramming systems and language workbenches (e.g., Rascal [35], TXL [10, 11], ANTLR [7], Stratego/XT [8], or Spoofax [31]) and language definition or executable semantic frameworks (e.g., K semantic framework [48] or PLT Redex [20]) as well as modeling frameworks (e.g., AM3 [1]). Further examples of SLRs include the repositories for Krishnamurthi's textbook on programming languages [37], Batory's Prolog-based work on teaching MDE [3], and Zaytsev et al.'s software language processing suite (SLPS) [52].

In this paper, we present an approach to testing and building SLRs with these overall characteristics:

**Relationship maintenance** The focus is on maintaining relationships across a heterogeneous collection of artifacts in a repository—this is a form of regression testing. Build management is supported in so far that relationships for function application can be selectively executed to derive missing or outdated artifacts. Package management [15, 12] and configuration management [16, 13] are not addressed.

**Languages as types** Artifacts in an SLR are 'typed' by languages. Languages are structured names typically associated with an algorithmic membership test. Language processing functionality (e.g., interpreters, conformance checkers, or transformations) is also 'typed' by languages. In this manner, artifacts can be systematically checked and the correct application of functionality can be systematically enforced.

**Heterogeneous representation** Multiple representations may co-exist for a language, as different metalanguages (or metaprogramming systems) and formats for serialization and interchange may be exercised. Representation types become part of the language names, as in *java(text)* versus *java(json)*: the former for Java's concrete syntactical, text-based representation; the latter for Java's abstract syntactical, JSON-based representation.

**Declaration of relationships** Testing and building are *not* described by scripting or a rule-based system. Instead, one declares relationships (e.g., for conformance) on artifacts based on the application of language processing functionality. Thus, testing and building basically requires checking and (re-) establishing declared relationships, thereby giving rise to a very simple and declarative semantics.

**Integrated compile- and run-time** The actual process of testing and building potentially involves compilation and execution across multiple programming languages and systems. By providing an integrated compile- and run-time, the process can be centrally administered. This language integration leverages command-line and foreign-function interfaces for invoking language processing functionality on files.

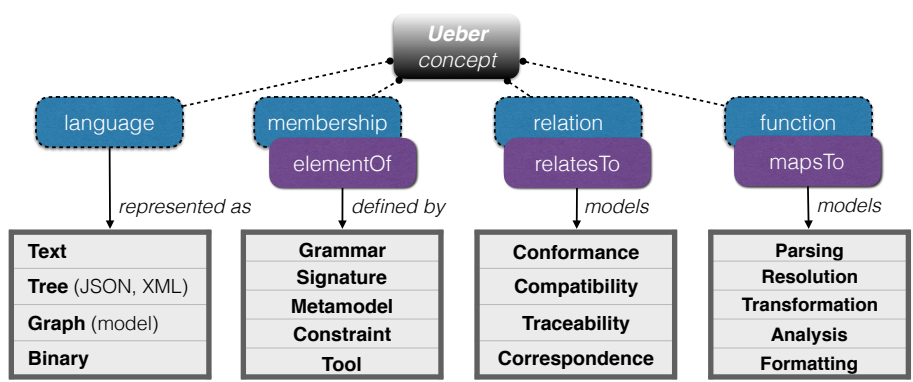

**Figure 1** Language concepts of Ueber.

**Contribution of the paper**  This paper describes a language design for testing and building SLRs—in accordance with the characteristics described above. The language design is realized in the domain-specific language Ueber. The language concepts are summarized in Figure 1.

That is, languages may be declared (concept 'language'); they may be associated with membership tests, e.g., based on grammar-based parsing (concept 'membership'), and artifacts may be typed by languages (concept 'elementOf'). Relations and functions on languages can be registered as plugins (concepts 'relation' and 'function') and they may be applied to artifacts (concepts 'relatesTo' and 'mapsTo'). At the bottom of the figure, we identify different representation types for languages (e.g., text, JSON, and XML), different forms of defining membership tests (e.g., grammars and metamodels), some common forms of relations (e.g., conformance and correspondence) and functions (e.g., parsing and transformation).

The Ueber language with its underlying approach to testing and building is applied to a particular SLR: YAS—*Y*et *A*nother *S*LR.[1] YAS targets teaching and research on the foundations and engineering of software languages; it uses Haskell, Prolog, Java, and Python for implementing language processing functionality; YAS also exercises various other technologies for language processing, e.g., the ANTLR parser generator and the StringTemplate library for template processing. YAS is the codebase underlying the introductory textbook on software languages by this author.[2]

**Road-map of the paper**  Section 2 sketches YAS in terms of modeled languages, implementation languages, and language processing components. Section 3 describes the Ueber language by means of examples. Section 4 sketches the semantics of Ueber informally. Section 5 summarizes the executable language definition of Ueber; the dynamic semantics directly supports testing and building. Section 6 sketches Ueber's integrated compile- and run-time for compiling and executing functionality in different languages. Section 7 introduces an abstraction mechanism in Ueber: relationship patterns. Section 8 discusses related work. Section 9 concludes the paper.

---

[1] YAS website: http://www.softlang.org/yas
[2] The book's website: http://www.softlang.org/book

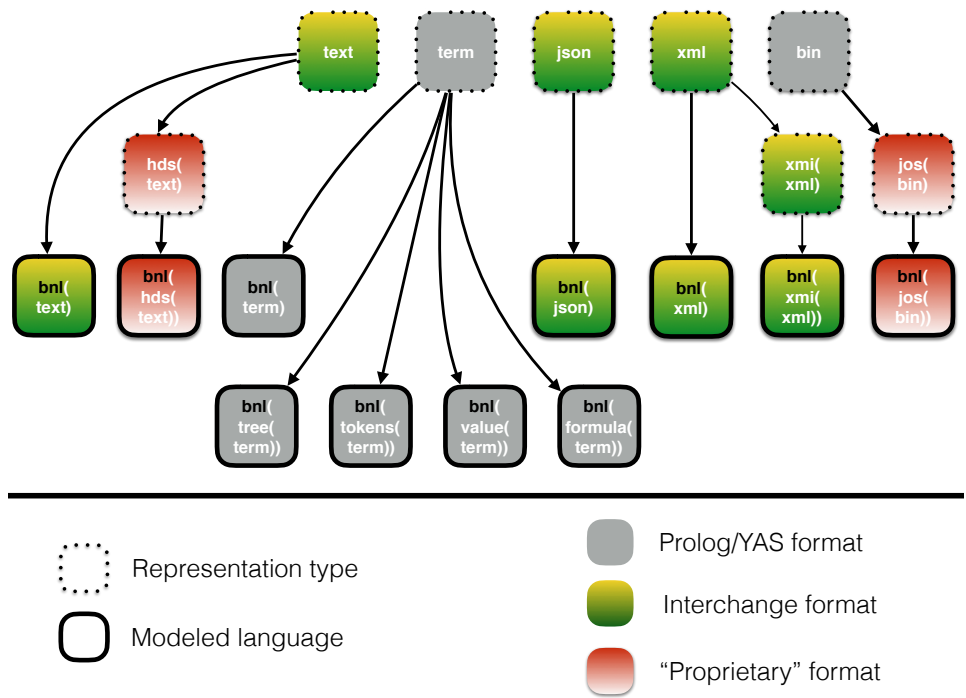

**Figure 2** An excerpt of YAS' language hierarchy: the representation types at the top (*text*, *term*, *json*, *xml*, *bin*) are 'universes' for text-, (Prolog) term-, JSON-, XML-based, or binary languages from which to draw languages as subsets. All the leafs of the subset hierarchy are BNL-related languages. There are some inner nodes (*xmi*(*xml*), *hds*(*text*), and *jos*(*bin*)); they correspond to language-independent formats for XMI, Java object serialization, and Haskell data serialization.

## 2  The YAS software language repository

### 2.1  Examples of languages

Figure 2 shows basic representation types in YAS and a few more specific software languages related to different aspects of a simple language BNL—*B*inary *N*umber *L*anguage. The nodes in the figure denote languages including 'formats' (e.g., XML-based ones) or general 'representation types' (e.g., *text*). The directed edges (arrows) denote subset relationship for languages in a set-theoretical sense. For instance, language *bnl*(*text*) corresponds to the concrete textual syntax of BNL. Thus, language *text* can be viewed as the universe for text-based languages. We explain the various languages in the sequel.

Here is an example of a binary number represented as text, i.e., an element of *bnl*(*text*):

Text resource *languages/BNL/samples/5comma25.bnl*

101.01

Language *bnl*(*json*) corresponds to the abstract, tree-based syntax of BNL using JSON for representation; here is the JSON representation of '101.01':

JSON resource *languages/BNL/samples/5comma25.json*

```
{
  "bits": ["one", "zero", "one"],
  "rest": ["zero", "one"]
}
```

Language *bnl*(*term*) corresponds to the abstract, tree-based syntax of BNL using Prolog terms for representation; here is '101.01' once represented as a prefix term:

Term resource *languages/BNL/samples/5comma25.term*

```
number(
  many(one, many(zero, single(one))),
  rational(many(zero, single(one))) ).
```

Language *bnl*(*tree*(*term*)) corresponds to the format for concrete syntax trees (CSTs) for BNL using again terms for representation. (We omit an example here because CSTs are verbose.) Language *bnl*(*tokens*(*term*)) corresponds to the representation format for tokenized binary numbers; here is '101.01' in tokenized form:

Term resource *languages/BNL/samples/5comma25.tokens*

```
['1', '0', '1', '.', '0', '1'].
```

Language *bnl*(*value*(*term*)) corresponds to the representation format for results when converting binary to decimal numbers. Here is the decimal value of '101.01':

Term resource *languages/BNL/samples/5comma25.value*

```
5.25.
```

Language *bnl*(*formula*(*term*)) corresponds to the representation format for formulae denoting the symbolic conversion of binary to decimal numbers. Here is the formula for '101.01':

Term resource *languages/BNL/samples/5comma25.formula*

```
2^ (1+1+1—1)+ (0+2^ (1+1+1—1—1))+ (0+2^ (—1—1)).
```

These are the remaining languages in Figure 2:

| | |
|---|---|
| **xml** | The universe for XML-based representation formats. |
| **xmi(xml)** | The XML-based representation format for EMF models. |
| **bin** | The universe for binary representation formats. |
| **jos(bin)** | The binary serialization format for Java objects. |
| **hds(text)** | The text-based serialization format for Haskell data. |
| **bnl(···)** | Representation formats for BNL as subsets of the aforementioned formats. |

XMI (named here *xmi(xml)*) is an established format in the space of model-driven engineering and metamodeling. JOS (named here *jos(bin)*) is our acronym for the format used by Java's basic approach to object serialization based on the interface *java.io.Serializable*.[3] HDS (named here *hds(text)*) is our acronym for the format used by Haskell's basic approach to data serialization based on the type classes *Read* and *Show*.[4]

---

[3] https://docs.oracle.com/javase/8/docs/technotes/guides/serialization/
[4] https://www.haskell.org/onlinereport/basic.html#sect6.3.3

**Relationship Maintenance in Software Language Repositories**

## 2.2 Examples of language processing components

There are three kinds of language processing components in YAS:

- Algorithmic membership tests for languages, e.g., based on parsing.
- Relations on languages for checking, for example, conformance.
- Functions on languages for computing artifacts, e.g., by transformation.

The following ANTLR-based grammar defines the language *bnl*(*text*), i.e., the text-based, concrete syntax of BNL; ANTLR's parser descriptions uses a grammar notation based on Extended Backus Naur Form (EBNF):

ANTLR resource *languages/BNL/ANTLR/org/softlang/BnlEbnf.g4*

```
grammar BnlEbnf;
@header { package org.softlang; }
number : bit+ ('.' bit+)? WS? EOF;
bit : '0' | '1';
WS : [ \t\n\r]+ ;
```

The following grammar also defines the language *bnl*(*text*), but this time in terms of YAS' grammar notation BGL—*B*asic *G*rammar *L*anguage, which is a variant of Backus Naur Form (BNF):

BGL resource *languages/BNL/cs.bgl*

```
[number] number : bits rest ;
[single] bits : bit ;
[many] bits : bit bits ;
[zero] bit : '0' ;
[one] bit : '1' ;
[integer] rest : ;
[rational] rest : '.' bits ;
```

The following signature defines the language *bnl*(*term*), i.e., the term-based (tree-based), abstract syntax of BNL; we use YAS' signature notation BSL—*B*asic *S*ignature *L*anguage:

BSL resource *languages/BNL/as.bsl*

```
symbol number: bits × rest → number ;
symbol single: bit → bits ;
symbol many: bit × bits → bits ;
symbol zero: → bit ;
symbol one: → bit ;
symbol integer: → rest ;
symbol rational: bits → rest ;
```

Figure 3 organizes language processors for BNL in a graph. The ellipsoid nodes are BNL-related languages, as discussed earlier. The rectangular nodes are language processing components; we only consider functions here; there are no relations. The edges identify the input and output types (i.e., languages) of the components. Let us briefly describe the components:

**scan**                                                       Map BNL text to BNL tokens
**parse**                                       Map BNL text or tokens to BNL terms

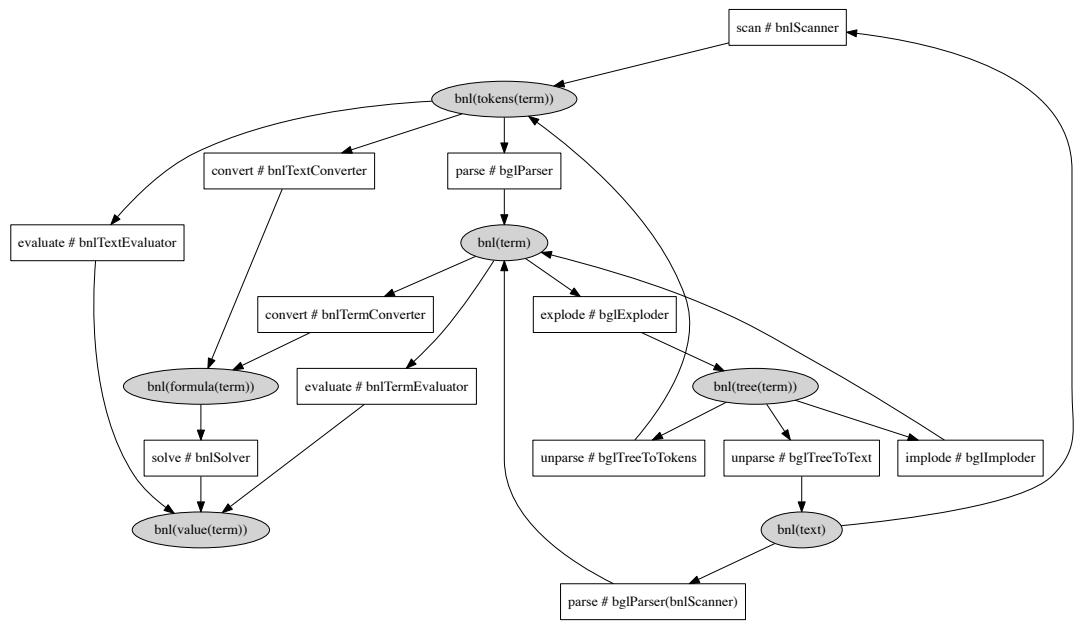

**Figure 3** An excerpt of YAS' language processing functionality.

| | |
|---|---|
| **unparse** | The inverse of parse |
| **explode** | Map BNL terms to BNL parse trees |
| **implode** | The inverse of explode |
| **evaluate** | Evaluate BNL text, tokens, or terms |
| **convert** | Convert BNL text, tokens, or terms to formulae |
| **solve** | Solve formulae to compute values |

For instance, the following Prolog module converts BNL tokens to formulae denoting the conversion of binary to decimal numbers. The module provides one implementation of the *convert* function noted above.

Prolog resource *languages/BNL/Prolog/bnlTextToFormula.pro*

```
number(Val1+Val2) ⟶ bits(Len1−1, Len1, Val1), rest(Val2).
bits(Pos, 1, Val) ⟶ bit(Pos, Val).
bits(Pos0, Len1+1, Val1+Val2) ⟶ bit(Pos0, Val1), bits(Pos0−1, Len1, Val2).
bit(_Pos, 0) ⟶ ['0'].
bit(Pos, 2^Pos) ⟶ ['1'].
rest(0) ⟶ [].
rest(Val) ⟶ ['.'], bits(−1, _Len, Val).
```

In fact, the module is programmed as a definite clause grammar (DCG) which represents Knuth's attribute grammar for number conversion from his seminal paper on attribute grammars [36]. The following module evaluates formulae as synthesized by the DCG:

Prolog resource *languages/BNL/Prolog/bnlEvaluator.pro*

```
evaluate(F, V) ⇐ V is F.
```

**Relationship Maintenance in Software Language Repositories**

## 2.3 YAS in numbers

An SLR may easily feature dozens if not hundreds of languages and functions. Testing and building such a repository as well as understanding problems during testing and building may be quite complex. Here is a short summary of YAS, as of writing:[5]

- 107 languages. (This includes different representation types.)
- 558 language-typed artifacts.
- 121 language-typed functions.
- 391 function applications.
- 252 Prolog modules.
- 171 Haskell modules.
- 111 Java classes.
- 19 Python scripts.

Let us convey the diversity of software languages and processing functionality in YAS. Here is a short list of languages that are defined in YAS or otherwise exercised in terms of language processing functionality:

**BNL**  *B*inary *N*umber *L*anguage.
**BIPL & EIPL**  *B*asic & *E*xtended Imperative Programming Language.
**BFPL & EFPL**  *B*asic & *E*xtended Functional Programming Language.
**FSML**  *F*inite *S*tate *M*achine *L*anguage
**BGL & EGL**  *B*asic & *E*xtended *G*rammar *L*anguage (variants of folklore BNF & EBNF).
**BSL & ESL**  *B*asic & *E*xtended *S*ignature *L*anguage (similar to algebraic data types).
**MML**  *M*eta*M*odeling *L*anguage (a variant of MOF, EMF).
**MMTL**  *M*eta*M*odeling *T*ransformation *L*anguage (refactorings et al. on metamodels).
**MMDL**  *M*eta*M*odeling *D*ifference *L*anguage (differences or deltas on metamodels).
**DDL**  Data Definition Language (as a subset of SQL).
**PPL**  Pretty Printer Language (combinators for formatting).
**TBL & GBL**  Tree- and graph-based Buddy Language (persons and their buddies).

## 3  UEBER in a nutshell

We define the abstract syntax of UEBER as an algebraic signature, i.e., the different UEBER constructs are modeled by algebraic constructor symbols. To this end, we use YAS' extended signature notation ESL—*E*xtended *S*ignature *L*anguage, which also incorporates support for type aliases and primitive types—in addition to just many-sorted constructor symbols.

---

[5] All Prolog-based components of YAS are fully managed with UEBER; many Haskell-, Java-, and Python-based still need to be adapted to be managed with the integrated compile- and run-time. This is planned for an upcoming YAS release.

ESL resource *languages/ueber/as.esl*

```
type model = decl* ;
symbol language : lang → decl ;
symbol elementOf : file × lang → decl ;
symbol notElementOf : file × lang → decl ;
symbol membership : lang × goal × file* → decl ;
symbol relation : rela × lang* × goal × file* → decl ;
symbol relatesTo : rela × file* → decl ;
symbol function : func × lang* × lang* × goal × file* → decl ;
symbol mapsTo : func × file* × file* → decl ;
symbol equivalence : lang × goal × file* → decl ;
symbol normalization : lang × goal × file* → decl ;
symbol macro : goal → decl ;
type file = string ; // filenames
type rela = string ; // names of relations
type func = string ; // names of functions
type lang = term ; // names of languages
type goal = term ; // Prolog literals
```

We go through the constructs, one by one. In the examples, we refer to artifacts from Section 2.

**Language declarations**   These introduce languages by name. For instance, the following declarations introduce the basic representation type *term* and a subset *bnl(term)* meant for the term-based representation of binary numbers.

```
language(term).
language(bnl(term)).
```

Thus, language names are terms; constants are used for basic representation types. When functors of arity 1 are applied to existing language names to form new language names, then a subset relationship is declared.

**Element declarations**   These assign languages to artifacts. For instance, the following declaration claims that the earlier BNL sample is indeed an element of the (text-based) BNL language:

```
elementOf('languages/BNL/samples/5comma25.bnl', bnl(text)).
```

In this manner, SLRs are organized as collections of language-typed artifacts. Arguably, any artifact in an SLR should be declared to be an element of at least one language.

**Membership declarations**   These make *elementOf* declarations checkable. For instance, a textual language should be associated with an acceptor (a parser); a term (tree) language should be associated with a conformance check for terms relative to a given signature. A membership declaration associates a language with a predicate to be applied to the content of the artifact to be verified. There are also negated *elementOf* declarations; see the *notElementOf/1* functor. These declarations are to be verified by failing membership tests.

For instance, the following declaration associates the text-based BNL language with a predicate which essentially interprets a context-free grammar:

**Relationship Maintenance in Software Language Repositories**

```
membership(bnl(text),
  bglTopDownAcceptor(bnlScanner), ['languages/BNL/cs.term']).
```

That is, the predicate *bglTopDownAcceptor/3* (with two positions readily filled in) is meant to check *bnl(text)* membership for artifacts. The first argument, `bnlScanner`, identifies Prolog functionality for scanning BNL. The second argument, `'languages/BNL/cs.term'`, refers to the grammar for the membership test; this argument is provided in a separate argument list of the declaration so that UEBER may make the given file path absolute.

**Relation & relatesTo declarations**  These declare language-typed relations and apply them to appropriately typed artifacts. UEBER relations have a name and are associated with a predicate of suitable arity. Here is a *relation* declaration for signature-based conformance and an illustrative *relatesTo* declaration for a conformance relationship:

```
relation(conformsTo, [term, bsl(term)], bslConformance, []).
relatesTo(conformsTo,
  ['languages/BNL/samples/5comma25.term',
   'languages/BNL/as.term']).
```

**Function & mapsTo declarations**  These declare language-typed functions and apply them to appropriately typed artifacts. If we were only interested in testing, relations would be sufficient. However, since we are also interested in building, functions add essential expressiveness, as their 'direction' may be operationally used by the UEBER semantics to derive missing or outdated assets.[6] Here is a *function* declaration for parsing text to terms for the BNL language and an illustrative *mapsTo* declaration to impose a corresponding relationship on artifacts:

```
function(parse,
  [bnl(text)], [bnl(term)],
  bglTopDownParser(bnlScanner), ['languages/BNL/cs.term']).
mapsTo(parse,
  ['languages/BNL/samples/5comma25.bnl'],
  ['languages/BNL/samples/5comma25.term']).
```

Arguably, no artifact in an SLR exists in isolation; thus, any given artifact in an SLR should be related to another artifact—by means of either a *relatesTo* or a *mapsTo* declaration.

We should also note that a name of a relation or a function may be used in multiple relation or function declarations and thus, it may be associated with multiple predicates, even with the same languages for inputs and outputs. This sort of overloading helps with organizing and testing alternative implementations which can be simply referred to by their 'shared' name. Thus, no effort is required to verify that alternative implementations agree with each other.

---

[6] In the rest of the paper, we assume that relations are treated like functions without outputs, i.e., *relation* and *relatesTo* declarations are translated to *function* and *mapsTo* declarations, respectively.

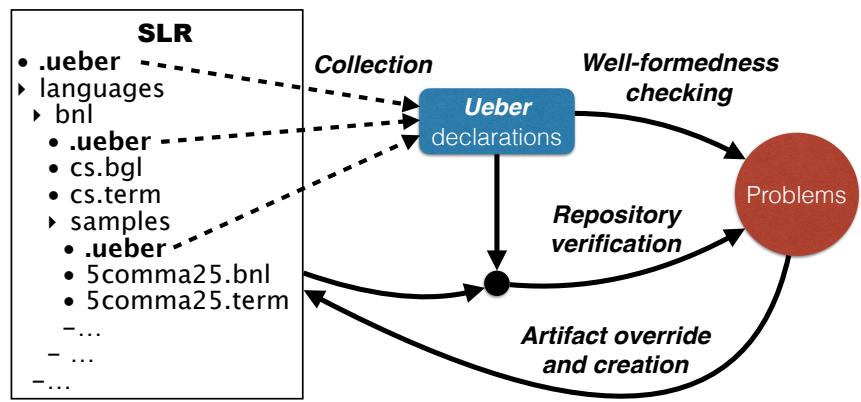

**Figure 4** Phases of Ueber processing.

**Equivalence & normalization declarations** These declarations associate additional functionality (predicates) with languages. A predicate for equivalence is applied whenever function outputs are compared with SLR baselines. A predicate for normalization is applied whenever function/relation inputs are read from the SLR. Equivalence and normalization is meant to make regression testing more robust, also in the view of alternative, possibly vacuously different implementations of relation and function names. For instance, sensitivity regarding formatting, naming, and order can be reduced in this manner.

**Macro declarations** These declarations facilitate instantiation of reusable patterns of Ueber declarations; this topic is deferred to Section 7.

## 4 Informal Ueber semantics

An executable, logic programming-based definition of Ueber is presented in Section 5. We explain the Ueber semantics here informally while also providing examples of testing and building-related problems that can be revealed and addressed with Ueber. Figure 4 shows the phases of processing Ueber declarations by its language implementation which is also part of YAS.

As illustrated by the dashed arrows on the left, Ueber declarations are typically distributed over the repository and stored in '.ueber' files to be close to the artifacts that should be constrained. Accordingly, a *collection* is performed to gather all Ueber declarations. Subsequently, *well-formedness checking* is applied to all declarations. One may think of well-formedness checking as the static semantics of Ueber. The guiding principle is to identify problems that do not involve yet access to the (other) files in the repository, thereby focusing on the integrity of the Ueber declarations as such. In particular, no predicates assigned by the declarations are applied yet. Eventually, a *repository verification* is performed based on the Ueber declarations. Thus, *elementOf*, *relatesTo*, and *mapsTo* relationships are evaluated on actual artifacts in the repository. One may think of verification as the dynamic semantics of Ueber.

**Relationship Maintenance in Software Language Repositories**

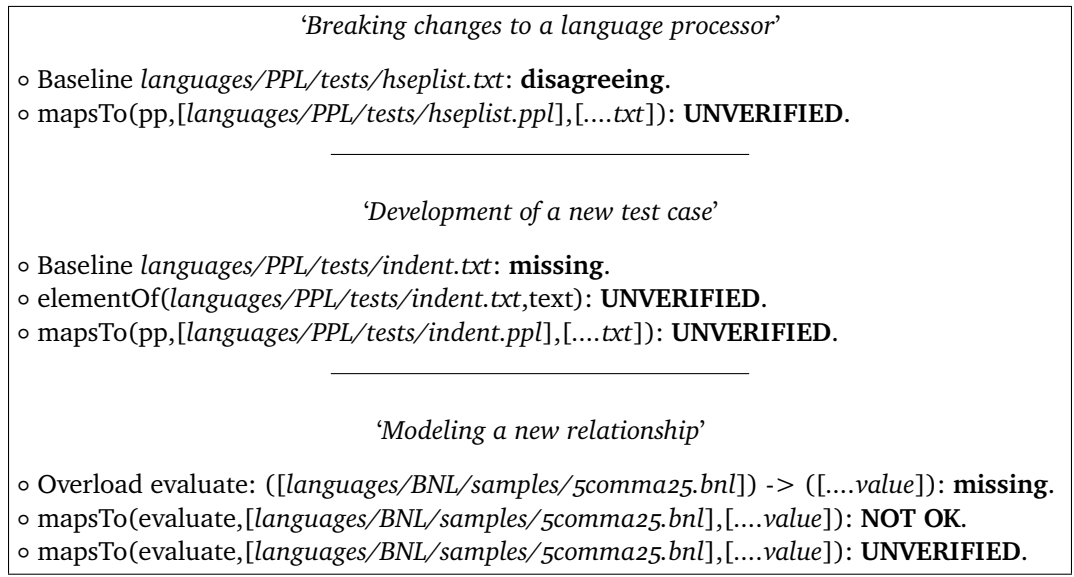

■ **Figure 5** Situations during language processor development with feedback provided by UEBER. (The '...' elisions hint at common filename prefixes in an error message.)

Both well-formedness checking and verification may produce 'problems'; these phases do not change the repository. A user of the UEBER processor may select though modi for overriding to update (apparently outdated) artifacts and creation to create (apparently missing) artifacts.

Let us look at three different situations ('problems'). The corresponding feedback of the UEBER implementation is summarized in Figure 5. We discuss the situations one by one:

**Breaking changes to a language processor** Let us assume that we are concerned with development (maintenance) of a pretty printing engine which processes 'box' expressions with operators for vertical and horizontal alignment of components. The engine processes box expressions according to PPL—YAS' *Pretty Printer Language*. The semantics of PPL expressions renders them as text. Test cases are captured as 'mapsTo' declarations from inputs (.ppl files) to outputs (.txt files)—the latter to be regarded as baselines. When the engine is 'broken', then test cases may fail as signaled in the figure. That is, the actual 'mapsTo' declaration is communicated as 'UNVERIFIED' and the relevant file is identified as a disagreeing baseline.

**Development of a new test case** We assume that a new test case is designed so that a certain box expression is rendered according to the semantics of PPL. In the beginning, the baseline may be missing. Rather than authoring the baseline explicitly, we may want to just render the new expression and inspect the result to see whether it can be captured as a suitable baseline. Until then, the file for the baseline is reported as missing and the corresponding declarations for *elementOf* and *mapsTo* relationships are reported as 'UNVERIFIED'.

**Modeling a new relationship** We assume that a new relationship between artifacts is discovered or suspected. For instance, we may assume that elements of *bnl(text)* (binary numbers) can be evaluated to return results as elements of *bnl(value(term))*.

Well-formedness checking of all declarations in the SLR finds that there is nowhere a suitable function declaration. Thus, the suggested relationship is already marked as 'NOT OK' (i.e., ill-formed), effectively implying that the relationship also ends up as 'UNVERIFIED'.

## 5  Executable definition of UEBER

We sketch some key aspects of UEBER's static and dynamic semantics. The complete definition is available online and readily linked from the module headers below.

### 5.1  Well-formedness checking

Well-formedness is modeled by the predicate *ok/2* with one clause per declaration form; the first argument serves as an environment with the complete set of declarations to be observed; we show the clause for *mapsTo* declarations as an example:

Prolog module *ueberOk.pro*

```
1  ...
2  ok(Ds, mapsTo(R, InFs, OutFs)) ⇐
3     assume(
4       member(function(R, _, _, _, _), Ds),
5       'Function ~w: missing.', [R] ),
6     map(ueberOk:assumeFile(Ds), InFs),
7     map(ueberOk:assumeFile(Ds), OutFs),
8     assume(
9       ueberDispatch:overloads(Ds, R, InFs, OutFs, [_|_]),
10      'Overload ~w:(~w) —> (~w): missing.', [R, InFs, OutFs] ).
11 ...
```

We use design-by-contract here: *assume(G, A, Ps)* is identical to *once(G)* except that in the case of *G*'s failure (*A, Ps*) is recorded as 'problem' in the knowledge base. A *mapsTo(R, InFs, OutFs)* declaration is Ok, if all of the following conditions hold: i) there is a function declaration for *R* (lines 5); ii) the filenames *InFs* and *OutFs* have some associated languages (see *assumeFile/2*; lines 6–7); iii) at least one overload of *R* is applicable to the given files in terms of their declared languages (line 9).

An interesting, non-trivial detail of well-formedness checking is overloading resolution for applying functions. An overload for a function is modeled as a pair consisting of a predicate *Pred* and arguments *Args*; these components correspond to the last two arguments of a function declaration. Given a function name *R* and filenames *InFs* and *OutFs*, the corresponding overloads are determined as follows:

Prolog module *ueberDispatch.pro*

```
1  overloads(Ds, R, InFs, OutFs, Overloads2) ⇐
2    findall( (Pred, Args, InLs, OutLs), (
3        member(function(R, InLs, OutLs, Pred, Args), Ds),
4        map(ueberDispatch:inferredLanguage(Ds), InFs, InLs),
5        map(ueberDispatch:inferredLanguage(Ds), OutFs, OutLs) ),
6      Overloads1 ),
7    findall( (Pred1, Args1), (
```

```
8        member((Pred1, Args1, InLs1, OutLs1), Overloads1),
9        \+ (
10           member((_, _, InLs2, OutLs2), Overloads1),
11           \+ (InLs1, OutLs1) == (InLs2, OutLs2),
12           map(languageTowardsBase, InLs2, InLs1),
13           map(languageTowardsBase, OutLs2, OutLs1) )),
14    Overloads2 ).
```

That is, all function declarations for *R* are filtered to determine those with suitable argument and result languages (lines 2–6). To this end, we seek languages *InLs* and *OutLs* such that the files *InFs* and *OutFs* are declared as elements of those languages or super-languages thereof. The helper predicate *inferredLanguage/3* lets us consider all super-languages of files. The rest of the clause (lines 7–14) further filters the overloads to favor more specific options (in terms of argument and result languages). The assumption is here that more specific options shadow more general options.

## 5.2 Repository verification

Verification is modeled by the predicate *verify/2* with one clause per declaration form; the first argument serves again as an environment with the complete set of declarations to be observed. Verification is only relevant for few declaration forms.

Prolog module *ueberVerify.pro*

```
1  verify(Ds, elementOf(F, L)) ⇐
2    ueberIO:readFile(F, L, Content1),
3    ueberNorm:normalize(Ds, F, L, Content1, Content2),
4    \+ (
5      languageTowardsBase(L, B),
6      member(membership(B, Pred, Args), Ds),
7      Pred =.. [Sym|_],
8      \+ assume(
9        ueberFFI:if(Sym, once(ueberFFI:invoke(Pred, Args, [L], [], [Content2], []))),
10       'File ~w element of language ~w according to ~w: failed.', [F, B, Pred] ) ).
11
12 verify(Ds, notElementOf(F, L)) ⇐ ...
13
14 verify(Ds, mapsTo(R, InFs, OutFs)) ⇐
15   ueberDispatch:overloads(Ds, R, InFs, OutFs, Overloads),
16   Overloads = [_|_],
17   map(ueberApply:apply(Ds, R, InFs, OutFs), Overloads).
```

An *elementOf(F, L)* declaration (lines 1–10; likewise for *notElementOf(F, L)*) is verified by trying out all applicable membership tests—these are membership tests associated with *L* as well as its super-languages. Any failing membership test is reported by using the *assume/3* predicate, just as in the case of well-formedness checking. A *mapsTo(R, InFs, OutFs)* declaration (lines 14–17) is verified by trying out the application of all applicable function overloads—they are determined as in the case of well-formedness checking with the predicate *overloads/5*.

An interesting, non-trivial detail of verification is the application of functions. Given a function name *R*, filenames *InFs* and *OutFs*, a predicate *Pred*, and arguments *Args*, function application is verified as follows:

Prolog module *ueberApply.pro*

```prolog
1  apply(Ds, R, InFs, OutFs, (Pred, Args)) ⇐
2    Pred =.. [Sym|_],
3    % Determine languages of files and read them in
4    map(ueberDispatch:declaredLanguage(Ds), InFs, InLs),
5    map(ueberDispatch:declaredLanguage(Ds), OutFs, OutLs),
6    map(ueberIO:readFile, InFs, InLs, InArgs1),
7    map(ueberIO:tryReadFile, OutFs, OutLs, Expected),
8    % Normalize function arguments
9    map(ueberNorm:normalize(Ds), InFs, InLs, InArgs1, InArgs2),
10   % Create variables for actual result
11   length(Expected, Len),
12   length(Actual, Len),
13   % Apply predicate
14   ueberFFI:if(
15     Sym,
16     (
17       assume(
18         once(ueberFFI:invoke(Pred, Args, InLs, OutLs, InArgs2, Actual)),
19         'Overload ~w#~w(~w)—>(~w): failed.',
20         [R, Pred, InFs, OutFs] ),
21       % Compare expected and actual results
22       map(ueberEq:compare(Ds), OutFs, OutLs, Expected, Actual)
23     )
24   ).
```

The files *InFs* and *OutFs* are read (lines 3–7); verification fails for a missing file in the former but not in the latter list (see *readFile/3* versus *tryReadFile/3*). Inputs are normalized (line 9). Fresh variables are prepared for the actual outputs (lines 11–12). The predicate is applied to the normalized content for the inputs and variables for the outputs (line 18). Upon successful application, expected and actual outputs are compared (line 22). The invocation of the predicate is shielded by a condition (see the use of *ueberFFI:if/2* in lines 14–...) to selectively deactivate foreign functions, as discussed in Section 6.

We omit the details of comparison; we only describe it informally here. Essentially, data needs to be compared for equality modulo equivalence and the baseline may also be missing:

- If expected and actual content are found to be non-equivalent, this would normally be reported as a regression issue. Subject to inspection, the developer may be able to confirm that the baseline is outdated and should be overridden. This behavior can be requested by the developer by running the process in *mode(override)*.

- If the expected content is missing (i.e., the corresponding file for the baseline could not be read), then this may mean that the developer has not yet authored a baseline. Subject again to inspection, the developer may be able to confirm that the computable content should be persisted as a baseline. This behavior can be requested by the developer by running the process in *mode(create)*.

**Relationship Maintenance in Software Language Repositories**

|  | Prolog | Haskell | Java | Python |
|---|---|---|---|---|
| **Implement relations & functions** | Predicates | Main functions | Main methods | Scripts |
| **Code location** | Module auto loading | Automated module search path | Automated CLASSPATH | Automated PYTHONPATH |
| **Compilation** | N/A | On the fly | On the fly | N/A |

■ **Figure 6** Ueber's integrated compile- and run-time.

|  | Prolog | Haskell | Java | Python |
|---|---|---|---|---|
| **Parse text** | DCG, … | Parsec, … | ANTLR, … | ANTLR, … |
| **Represent trees** | Prolog terms JSON, XML, … | read/show conversion, JSON, XML | JSON, XML | str/repr conversion, JSON, XML |
| **Represent graphs** | Prolog terms | N/A | Serializable objects, XMI, … | Serializable objects, … |

■ **Figure 7** Representation across different implementation languages.

## 6 Ueber's integrated compile- and run-time

The overall objective of an integrated compile- and run-time is that language processing functionality can be implemented in different languages and these implementations are compiled and executed as part of processing Ueber declarations in an SLR.

Figure 6 identifies some aspects of integration and how they apply to different implementation languages in YAS. That is, language processing functionality is obviously implemented as predicates in YAS' native Prolog language, while main functions are leveraged for Haskell, main methods for Java, and plain scripts for Python. The I/O data for applying functions and relations resides in language-typed files. The actual implementations of functions and relations do not operate, though, directly on these files, but they access temporary files (i.e., copies) or in-memory manifestations instead so that Ueber can effectively perform normalization, check on regression, and handle overriding or creation of artifacts, as necessary.

The location of code for language processing functionality relies on simple conventions: all Prolog modules in Ueber-managed folders are auto-load; all folders with Ueber-hosted references to Java classes are automatically added to the CLASSPATH; likewise for Python and Haskell. Compilation, where necessary (Java and Haskell), is performed on the fly before attempting execution of the referenced functionality.

Figure 7 hints at processing and representing artifacts of text- versus tree- versus graph-based languages with (in) the different implementation languages. That is, text-based languages are to be processed with whatever parsing approach is available; artifacts of tree-based languages may always be represented in XML or JSON, but some languages may provide some native conversion approach involving a text representation. For instance, in Haskell, one can effectively serialize terms of algebraic data types into strings, subject to the uniform implementation of the Read/Show type classes. Artifacts of graph-based languages require (object) serialization, e.g., Java's *java.io.Serializable* or EMF's XMI format.

Foreign functionality is integrated by UEBER declarations that use designated meta-predicates with symbols *haskell*, *java*, and *python* specifying the foreign language and an argument position for referring to FFI code units. For instance, the following declaration integrates a Java class (its main method) as an implementation of a membership test for *bnl*(*text*).

Ueber resource *languages/BNL/ANTLR/.ueber*

```
membership(bnl(text), java('org.softlang.BnlEbnfAcceptor'), [.])
```

(By convention, the FFI predicate receives an extra argument '.' to be aware of the current directory.) The actual functionality is implemented as follows:

Java resource *languages/BNL/ANTLR/org/softlang/BnlEbnfAcceptor.java*

```java
package org.softlang;

import org.antlr.v4.runtime.*;
import java.io.IOException;

public class BnlEbnfAcceptor {
  public static void main(String[] args) throws IOException {
    BnlEbnfParser parser =
      new BnlEbnfParser(
        new CommonTokenStream(
          new BnlEbnfLexer(
            new ANTLRFileStream(args[0]))));
    parser.number();
    System.exit(parser.getNumberOfSyntaxErrors());
  }
}
```

This code would be auto-compiled before it is executed. The return code is used to communicate success/failure of the membership test. Without going into low-level details of interacting with foreign functions, the relevant entity of the UEBER semantics is the *invoke/6* predicate which executes language processing functionality; the predicate can be viewed as the foreign function interface (FFI) of UEBER. Here is a sketch:

Prolog module *ueberFFI.pro*

```prolog
...
% Supported FFI languages
ffi_language(java).
ffi_language(python).
ffi_language(haskell).

% Test a predicate to be foreign
ffi_call(Pred) ⇐ Pred =.. [Sym|_], ffi_language(Sym).

% Invoke functionality
```

**Relationship Maintenance in Software Language Repositories**

```
invoke(Pred, Args, InLs, OutLs, InArgs, OutArgs) ⇐
    \+ ffi_call(Pred) —>
        % Apply Prolog predicate right away on content
        concat([Args, InArgs, OutArgs], AllArgs),
        apply(Pred, AllArgs)
    ;
    % Handle different FFI languages by CLI and temporary files
        ...
        .
...
```

The first branch of *invoke/6* shows that Prolog predicates would be simply applied on the content for input and output (if any). The second (and elided) branch of *invoke/6* would set up and check temporary files and call foreign functions through the command-line interface of the operating system in a language-specific manner.

## 7 Relationship patterns

An SLR may contain a large number of artifacts and there may be recurring patterns of groups of artifacts with associated relationships. UEBER provides an abstraction mechanism to capture such relationship patterns. In fact, patterns are modeled as predicates that generate UEBER declarations to be added to the knowledge base with a designated *ueber/1* predicate. Consider the following simple example of a pattern:

UeberProlog resource *languages/ueber/macros/fxy.pro*

```
fxy(Fun,FX,LX,FY,LY) ⇐
    ueber([
        elementOf(FX, LX),
        elementOf(FY, LY),
        mapsTo(Fun, [FX], [FY]) ]).
```

The *fxy/5* pattern combines a function application with *elementOf* declarations for argument and result. This is clearly a very common scenario. To see how patterns are applied (i.e., instantiated), consider the following UEBER declarations from a test suite for a language implementation. The UEBER construct *macro/1* is used to apply the pattern *fxy/5* several times:

Ueber resource *languages/PPL/tests/.ueber*

```
[
  macro(fxy(pp, 'text.ppl', ppl(term), 'text.txt', text)),
  macro(fxy(pp, 'vbox.ppl', ppl(term), 'vbox.txt', text)),
  macro(fxy(pp, 'vlist.ppl', ppl(term), 'vbox.txt', text)),
  ...
].
```

The shown declarations are concerned with PPL's 'box' expressions, as discussed earlier. The *fxy/5* pattern is used for each test case; see how the the .ppl files are mapped to the .txt files; the text files are the baselines.

Here is another example of a pattern which models parsing from a text-based to a term-based language. The pattern relies on the convention that a language's name is used as file extension.

UeberProlog resource *languages/ueber/macros/parse.pro*

```
parseFile(TextF) ⇐
  file_name_extension(Base, L, TextF),
  file_name_extension(Base, term, TermF),
  TextL =.. [L, text],
  TermL =.. [L, term],
  ueber([macro(fxy(parse, TextF, TextL, TermF, TermL))]).
```

That is, the language functor is extracted from the macro parameter for the input file for parsing, the text and term languages are composed, the filename for the output of parsing is assembled. Finally, the *fxy/5* pattern is used to assign types to input and output and to invoke the parse function. For instance:

```
macro(parseFile('languages/BNL/samples/5comma25.bnl')).
```

An important property of the *parseFile/1* pattern is that it abstracts from the realizations of the text and term languages; it only assumes their existence and a suitable parse function. The parse function is overloaded and the language memberships are implemented appropriately—in several different ways depending on the language at hand. That is, we may have used, for example, YAS' grammar languages BGL or EGL, or Prolog's DCGs, or a parser generator like ANTLR.

As an example of a more complex pattern, consider the following predicate supporting syntax definition based on BGL and BSL:

UeberProlog resource *languages/ueber/macros/bgl-and-bsl.pro*

```
basicSyntax(L) ⇐
  % Languages for representation
  TextL =.. [L, text],
  TokensL =.. [L, tokens(term)],
  TreeL =.. [L, bcl(term)],
  TermL =.. [L, term],
  % Synthesize scanner predicate
  atom_concat(L, 'Scanner', S),
  ueber([
    language(TextL),
    language(TokensL),
    language(TreeL),
    language(TermL),
    % Syntax definition artifacts
    macro(fxy(parse, 'cs.bgl', bgl(text), 'cs.term', bgl(term))),
    macro(fxy(parse, 'as.bsl', bsl(text), 'as.term', bsl(term))),
    macro(fxy(project, 'cs.term', bgl(term), 'as.term', bsl(term))),
    % Membership tests for artifacts
    membership(TextL, bglAcceptor(S), ['cs.term']),
    membership(TokensL, bglAcceptor, ['cs.term']),
    membership(TreeL, bclOk:main, ['cs.term']),
    membership(TermL, bslTerm, ['as.term']),
    % Functions for scanning, parsing, unparsing, etc.
    function(scan, [TextL], [TokensL], S, []),
    function(parse, [TokensL], [TermL], bglParser, ['cs.term']),
    function(parse, [TextL], [TermL], bglParser(S), ['cs.term']),
    function(cstToAst, [TreeL], [TermL], cstToAst, []),
    function(astToCst, [TermL], [TreeL], astToCst, ['cs.term']),
    function(unparse, [TreeL], [TokensL], bglTreeToTokens, []),
    function(unparse, [TreeL], [TextL], bglTreeToText, []) ]).
```

**Relationship Maintenance in Software Language Repositories**

That is, the *basicSyntax/1* pattern assumes that the basic notations BGL and BSL are used for concrete and abstract syntax definition and that text is to be scanned into tokens, further parsed into parse trees, imploded to terms, and likewise on the way back. If any of these many steps breaks for any language in the SLR, the assumed style of exposing various intermediate representations and maintaining their baselines helps with testing and maintenance.

The following collection of patterns facilitates testing of language implementations in general:

UeberProlog resource *languages/ueber/macros/test.pro*

```
% Positive sample for parsing
parseable(TextF) ⇐
  file_name_extension(_, L, TextF),
  TextL =.. [L, text],
  ueber([elementOf(TextF, TextL)]).

% Negative sample for parsing
unparseable(TextF) ⇐
  file_name_extension(_, L, TextF),
  TextL =.. [L, text],
  ueber([
    elementOf(TextF, text),
    notElementOf(TextF, TextL) ]).

% Positive sample for well−formedness
well_formed(TextF) ⇐
  file_name_extension(Base, L, TextF),
  file_name_extension(Base, term, TermF),
  TermL =.. [L, term],
  ueber([
    macro(parseFile(TextF)),
    elementOf(TermF, ok(TermL)) ]).

% Negative sample for well−formedness
ill_formed(TextF) ⇐
  file_name_extension(Base, L, TextF),
  file_name_extension(Base, term, TermF),
  TermL =.. [L, term],
  ueber([
    macro(parseFile(TextF)),
    notElementOf(TermF, ok(TermL)) ]).
```

The overall assumption underlying these macros is that one may have both positive and negative samples in terms of both syntax and well-formedness. The language functor is again extracted from the extension of the filename parameter; the corresponding text and term languages are constructed from the functor. The additional functor *ok/1* is applied on top of the term language to construct the language with well-formedness enforced. The patterns for negative samples leverage negated *elementOf* declarations, but only after they established that the given artifact is element of a suitable super-set.

YAS features further patterns that deal with graph-based representations, meta-modeling, term rewriting, and others.

## 8 Related work

The following areas of related work are identified: i) existing approaches to testing language processors; ii) declarative approaches to build management overall; iii) model management or megamodeling; iv) compile- and run-time integration.

### 8.1 Testing language processors

Language processors are typically tested in a manner inspired by unit testing. For instance, the transformation language *Stratego/XT* [8] incorporates a unit testing framework *StrategoUnit* or *SUnit* which is directly inspired by JUnit in the Java platform; there are test suites consisting of test cases; each test case applies some transformation to an input stating also the expected output including the option of failure. The executable semantic framework *K* [48] also incorporates a unit testing tool, *ktest*, which is quite versatile in terms of invoking different phases of executing semantics in different configurations, also subject to handling the rich algebraic structure (e.g., maps) of configurations. Several systems also support test-data generation, e.g., the executable semantic framework *PLT Redex* [20] with support for randomized testing [33] or the *Maude* language and system for equational and rewriting logic specification [42] with support for test-case generation from semantics [47] also based on more general techniques for test-case generation, code coverage, property-based testing, and conformance testing [45, 46]. Testing language implementations developed with a language workbench requires an additional element. That is, the IDE, e.g., in terms of produced warnings or errors, the content in specific views such as a package explorer, and the availability and the results of applying a refactoring, must be tested. The *SPoofax Testing language* (SPT) [29, 30] supports such IDE testing for languages implemented with the language workbench *Spoofax* [31].

The Ueber approach covers unit testing; in future work, it could and should be extended to cover test-data generation, possibly based on our previous work [38, 21] or the work cited above [47, 33]. Test-data generation has not been an urgency in YAS because the language processors are simple and small, as they target teaching. The Ueber approach does not readily address IDE testing. None of the cited approaches address the more repository-related characteristics of Ueber and its use in YAS in a heterogeneous setup: a) organization of languages in a subset-ordered nominal hierarchy; b) repository-scope aggregation and checking of declarations; c) overloading of language-typed, named relations and functions to permit alternative implementations; d) integrated compile- and run-time for multiple implementation languages; e) incorporation of equivalence and normalization.

### 8.2 Declarative build management

Language processors are typically built with a conservative rule-based build system such as Make or ant—just like any software. Arguably, rule-based build systems are declarative in nature, except that they may insufficiently track dependencies so that incremental building and testing may fail to be sound or optimal. A declarative approach to build management and configuration management has been proposed by Singleton and collaborators [50, 9]; in particular, logic deduction (Prolog) is used to represent Make-like rules; soundness and optimality of incremental building is not addressed. The Haskell-based build tool Shake [44] incorporates dependencies that are retrieved while building. In this manner, Shake better enables incremental builds. The Java API pluto [17] also addresses incremental builds; it comes with a formalization which shows that pluto provides a sound and optimal incremental build system with dynamic dependencies. A notion of safeness has also been studied for Make-based build management with coverage of incremental compilation [27].

Importantly, UEBER does not commit to a rule-based approach. UEBER and its application to YAS assumes a graph-based view on a repository where the artifacts are the nodes in the graph and the applications of language-processing functionality are essentially the edges in the graph, as they relate the nodes. When the relationships are applications of functions (as opposed to relations), then these 'directed' edges can also be executed for deriving missing or outdated artifacts. In this manner, testing and building are unified in our approach.

### 8.3 Model management

The notion of megamodeling or model management is due to Bézivin and collaborators [6, 1] and it was conceived in the context of the MDE (Model-driven engineering) technological space [5]. A megamodel provides a macroscopic view on a collection of MDE artifacts such as models, metamodels, and model transformations so that their relationships are understood or managed. One may say that our research applies the idea of megamodeling to testing and building software language repositories with the involvement of different technological spaces.

A megamodeling language related to UEBER is MoScript [34], as it realizes megamodeling in an executable manner for the purpose of querying and manipulating model repositories. MoScript provides operations such as *save* for saving a model in the repo, *remove* for removing a model from the repo, *allContentsInstances* (parameterized by a type) for querying all models of a given type, and *applyTo* for applying a transformation to models. There are also scripting-related concepts, e.g., binding variables or performing iteration ('for'). By contrast, UEBER is not a scripting language; language memberships of artifacts and relationships between artifacts are just declared. Testing and building is a consequence of the UEBER semantics.

Megamodeling is used in various contexts of software engineering. For instance, megamodeling in combination with model weaving can be used for inter-DSL coordination [28]. Megamodeling has also be shown to be helpful in describing the design space of options in parsing and unparsing and related phases [51]. Megamodeling has been applied too in the context of software process line modeling and evolution [49]. In our previous work on linguistic architecture of software systems [19, 39, 40], we used megamodels as abstractions over actual software systems; testing and building these systems was not a concern, as the systems themselves were always supposed to take care of these aspects already. A megamodeling-like technique has been used in a workbench for coupled model and metamodel evolution [25].

### 8.4 Compile- and run-time integration

Technically, Ueber's FFI is relatively straightforward; it is comparable to other domain-specific software which requires relatively standardized plugins in foreign languages; see, e.g., [14]. We assume that language-processing functionality is pure and strict and thus, no challenges of language composition arise [2]. Typing [23, 41] is also relatively straightforward because we only consider very few basic representation types for foreign functions (text, JSON, XML, bin). The Ueber approach relates to MoScript [34] (mentioned above) with its import/export mechanism and Rascal [26] with its notion of resources. However, our research shows that a small set of basic representation types can be used in combination with a command line-based interface for integrating language processing functionality across quite different implementation languages. Existing build and test systems could be possibly integrated by means of declarative wrappers, as it has been studied in the context of scripting [4].

## 9 Concluding remarks

**Summary** We have described a declarative approach to regression testing and build management for software language repositories (SLRs). The core contribution is a domain-specific, declarative language Ueber which manages the consistency of artifacts typed by languages with respect to language processing functionality again typed by languages. We have applied this approach to the SLR YAS with a Prolog-based implementation of Ueber that covers distribution of declarations over the repository, abstraction over relationship patterns, well-formedness checking, repository verification including aspects of testing without any side effects as well as building with side effects to the repository. While Ueber is implemented in Prolog and YAS uses Prolog for a significant part of its language processors, the overall approach and the particular Ueber implementation are not limited to Prolog, as we have shown by an integrated compile- and run-time for also Haskell, Java, and Python, subject to a relatively simple foreign function interface.

**Future work** The current state of YAS and Ueber suggests more research on declarative build management and regression testing as well as megamodeling in the broader context of software development or the more narrow context of software language repositories. We suggest several directions for future work.

*Incremental building*. Ueber's implementation, as it stands, performs a whole-repository analysis for checking and verification. For the current state of YAS, with just a few hundreds of tracked artifacts, the analysis takes about 15 seconds on a MacBook Air (i7, 2.2Ghz, 8GB, SWI-Prolog 7.2.3 default settings). An SLR size larger by one or two orders of magnitude is conceivable in which case an incremental approach is mandatory—certainly for interactive development. Incremental approaches are used in practice and soundness has been studied [17]. An SLR like YAS poses a challenge in so far that the metametalevel (thus, bootstrapping) is involved.

*Test-data generation.* In YAS, as it stands, testing relies on authored test cases listed as U<small>EBER</small> declarations. Test-data generators at the level of abstract and concrete syntax with and without extra semantic constraints could be integrated into the SLR approach. For instance, the coverage criteria, the declarative test-data generation approach, and language-oriented applications of our previous work [38, 21] may be leveraged in this context.

*Version control integration.* SLR and version control could be integrated in a mutually beneficial manner. In particular, some forms of validity checking could be automated. For instance, every artifact modeled by the SLR should eventually also be persisted via version control. In fact, any artifact under version control should also be modeled by the SLR.

*Linked Data.* An SLR—because of its complex structure with diverse types of entities and relationships—can benefit from extra means of helping with exploration and querying as well as connection with external knowledge resources. The *Linked Data* paradigm may be of use here; it has been applied in the related context of software chrestomathies [18].

**Acknowledgment**   Much of what is developed in the paper benefitted from collaboration with colleagues and several of my research students. I would like to thank, in particular, Jean-Marie Favre, Vadim Zaytsev, Andrei Varanovich, Lukas Härtel, Johannes Härtel, and Marcel Heinz.